\documentclass{svjour3}

\smartqed
\usepackage{amsmath,amssymb}
\usepackage{graphicx}
\usepackage{booktabs}
\usepackage[hidelinks]{hyperref}

\newcommand{\dd}{\mathrm{d}}
\newcommand{\vX}{\mathbf{X}}
\newcommand{\vY}{\mathbf{Y}}
\newcommand{\vz}{\hat{\mathbf{z}}}
\newcommand{\rhat}{\hat{\mathbf{r}}}
\newcommand{\ELV}{\mathcal{E}_{\mathrm{LV}}}
\newcommand{\half}{\tfrac12}

\hypersetup{
  pdftitle={Zero expansion is not free: the sharp minimum of Eulerian negative energy in Natario warp drives},
  pdfauthor={N. Bolivar, I. Vasilev, G. Abellan}
}

\journalname{General Relativity and Gravitation}

\begin{document}

\title{Zero expansion is not free: the sharp minimum of Eulerian
negative energy in Nat\'ario warp drives}

\titlerunning{The sharp minimum of Eulerian negative energy in Nat\'ario warp drives}

\author{N.~Bol\'ivar \and I.~Vasilev \and G.~Abell\'an}

\institute{
N.~Bol\'ivar \and G.~Abell\'an \at
Departamento de F\'isica, Facultad de Ciencias, Universidad
Central de Venezuela, Av.~Los Ilustres, Caracas 1041-A, Venezuela
\and
N.~Bol\'ivar \and I.~Vasilev \and G.~Abell\'an \at
Astrum Drive Technologies, Dallas Pkwy Unit 120 B, Frisco,
TX 75034, USA \\
\email{nelson.e.bolivar@ucv.ve}
}

\date{Received: date / Accepted: date}

\maketitle

\begin{abstract}
Nat\'ario's warp drive replaces the contraction and expansion of the
Alcubierre bubble by a divergence-free shift. We determine exactly how
much Eulerian negative energy this construction requires. For the
Lobo--Visser volume integral
$\ELV=-\int_{\Sigma_t}\rho_{\rm E}\,\dd^{3}x$ on the standard flat slice,
with unit lapse, a flat interior of radius $R$, and exact matching to a
uniform exterior stream of speed $v$ at $R+\Delta$, we prove that every
admissible shift obeys
$\ELV\ge \tfrac{1}{60}\,v^{2}R^{4}/\Delta^{3}\,
[(1+\Delta/2R)(1+\Delta/R)^{2}]^{-1}$,
and we find the exact minimum in closed form at every aspect ratio. As
the wall thins, $\mathcal E_{\min}\sim v^{2}R^{4}/(4\Delta^{3})$ with
sharp constant $1/4$, two powers of $R/\Delta$ above the estimate
$v^{2}R^{2}/\Delta$ applied to both the Nat\'ario and the Alcubierre
bubbles. The mechanism is kinematic. Incompressibility forces the flux
displaced by the bubble to return through the wall, and the tangential
shear of this return current dominates the energy. An exact flux
identity establishes the return current for every admissible shift,
without symmetry assumptions. The full variational problem is equivalent
to classical Stokes flow between concentric spheres. Its unique
minimizer combines $1,r^{2},r^{-1},r^{-3}$ and tends to the
minimal-curvature cubic $f^{*}(x)=\tfrac32x^{2}-x^{3}$ in the thin-wall
limit. Symbolic identities and independent three-dimensional quadrature
confirm each analytic step. The quantity bounded is a slice- and
observer-dependent measure of energy-condition violation, distinct from
the ADM mass.

\end{abstract}

\section{Introduction}
\label{sec:intro}

The Alcubierre warp drive~\cite{Alcubierre:1994tu} carries a flat region
of spacetime forward by contracting space ahead of it and expanding
space behind, at the cost of matter that violates the classical energy
conditions~\cite{Alcubierre:1994tu,Pfenning:1997wh,LoboVisser:2004}.
Nat\'ario~\cite{Natario:2001} observed that the contraction and
expansion are not essential: if one takes the Arnowitt--Deser--Misner
shift vector to be \emph{divergence-free}, the volume elements of the
Eulerian foliation are exactly preserved, and the bubble slides through
space rather than surfing on it. This ``warp drive with zero expansion''
is now one of the two standard warp-drive templates and underlies much
of the modern
literature~\cite{LoboVisser:2004,Santiago:2021,BobrickMartire:2021,Fuchs:2024,Schuster:2023,Rodal:2024}.
Here we determine exactly how much Eulerian negative energy a
zero-expansion bubble with a compactly matched spherical wall requires.
We prove a lower bound valid for every admissible shift and find the
exact minimum, whose thin-wall leading term is
$v^{2}R^{4}/(4\Delta^{3})$.

How large is the negative-energy volume integral on the standard
Eulerian slice? For the Alcubierre drive, Lobo and
Visser~\cite{LoboVisser:2004} gave a clean and widely used estimate.
Integrating the exact Eulerian energy
density over a wall of radius $R$ and thickness $\Delta=1/\sigma$, they
found
\begin{equation}
M_{\rm warp} \;\approx\; -\,v^{2}R^{2}\sigma
            \;=\; -\,\frac{v^{2}R^{2}}{\Delta}\,.
\label{eq:LV}
\end{equation}
They used this volume integral to compare the warp field with a ship
mass. For the Nat\'ario drive they noted that
``gradients of [the shift] in the bubble walls are of order $v\sigma$''
and concluded that the same estimate~\eqref{eq:LV} applies ``as for the
Alcubierre warp bubble''~\cite[Eqs.~(21)--(24)]{LoboVisser:2004}. This is
an entirely natural estimate---the radial part of the canonical shift varies by
$O(v)$ across a wall of width $\Delta$, giving gradients
$O(v/\Delta)$---but it does not control every component of a
divergence-free field.

The zero-expansion drive has an additional shear contribution that is
not captured by the estimate above. A divergence-free field cannot switch off abruptly:
whatever flows in from outside must go somewhere. In the frame comoving
with the bubble the exterior stream carries a volume flux $\pi v R^{2}$
toward the wall, and incompressibility forces all of it to pass
\emph{through} the wall, a shell of thickness only $\Delta$. For a fixed
rescaled canonical profile, the tangential shift inside the wall is of
order $vR/\Delta$ and its gradients are of order $vR/\Delta^{2}$. This
``return current'' produces shear absent from the diagonal part of the
extrinsic curvature. It dominates the thin-wall energy for this profile
family; the flux argument below gives a lower bound for the whole
admissible class.

The local density formula is implicit in Nat\'ario's strain tensor and
is written out explicitly by Rodal~\cite{Natario:2001,Rodal:2024}, who
also identifies the large second-derivative contribution. What has been
missing is control of its volume integral over the whole admissible
class of shifts with prescribed boundary data. We provide that control,
and we find that the extremal problem coincides with the classical
concentric-sphere Stokes resistance~\cite{Nangia:2017}. The
correspondence works in both directions. It gives warp-drive energetics
a closed-form minimizer, and it gives the classical Stokes solution a
relativistic meaning as the least Eulerian negative energy compatible
with the boundary data.

The quantity bounded here has a precise meaning. It is the integrated
negative energy density measured by the Eulerian congruence on the
standard slice, the quantifier introduced by Lobo and Visser and used
throughout the warp-drive literature to compare configurations. It is
distinct from the ADM mass, which vanishes here because the induced
spatial metric in Eq.~\eqref{eq:metric} is exactly Euclidean, in line
with the distinction among warp-drive mass notions emphasized in
Ref.~\cite{Schuster:2023}. It is also not a construction cost or a
statement about a quantum stress tensor. Within a fixed and widely used
convention, it measures the amount of local weak-energy-condition
violation seen by the Eulerian observers.

Our results are the following.
\begin{itemize}
\item[(i)] \emph{Mechanism} (Secs.~\ref{sec:setup}--\ref{sec:where}).
For Nat\'ario's canonical field the exact energy splits into the
radial-compression term captured by Eq.~\eqref{eq:LV} and a
return-current term larger by $(R/\Delta)^{2}$. A single explicit
example gives $1.97$ versus $435.4$.
\item[(ii)] \emph{Exact flux identity} (Sec.~\ref{sec:flux}). Every
admissible shift carries the displaced flux through the wall, with mean
speed $\simeq vR/2\Delta$ at the equator, and no symmetry is assumed.
\item[(iii)] \emph{Universal lower bound} (Sec.~\ref{sec:bound}).
Equation~\eqref{eq:bound} holds for every admissible shift with an
explicit constant, and its derivation loses no Korn constant.
\item[(iv)] \emph{Exact minimum} (Sec.~\ref{sec:optimal}). The
variational problem is solved in closed form at every aspect ratio
through its equivalence with Stokes flow. The sharp thin-wall constant
is $1/4$, attained by the minimal-curvature cubic.
\end{itemize}
Section~\ref{sec:numerics} checks each step against an independent
evaluation of the constraint, and Sec.~\ref{sec:discussion} draws the
physical consequences.

Throughout we use geometric units $G=c=1$; Latin indices are spatial;
and $\sigma_{ij}(\vX)=\half(\partial_i X_j+\partial_j X_i)$ is the strain
of a field $\vX$ on flat $\mathbb{R}^{3}$.

\section{The energy density, in closed form}
\label{sec:setup}

We consider the stationary, zero-expansion subclass of Nat\'ario's
metrics~\cite{Natario:2001},
\begin{equation}
\dd s^{2} = -\dd t^{2}
 + \delta_{ij}\,(\dd x^{i}-X^{i}\dd t)(\dd x^{j}-X^{j}\dd t),
\qquad \nabla\!\cdot\!\vX = 0,
\label{eq:metric}
\end{equation}
with a time-independent, divergence-free field $\vX$. In the ADM
convention $\dd x^i+\beta^i\dd t$, $\beta^i=-X^i$; we call $\vX$ the
shift field throughout. The future unit normal is
$n=\partial_t+X^i\partial_i$. The Eulerian slices
are intrinsically flat, the lapse is unity, and the extrinsic curvature
is $K_{ij}=-\sigma_{ij}(\vX)$; it is trace-free precisely because
$\vX$ is divergence-free, so the expansion
$\vartheta=\nabla_\mu n^\mu=\nabla\!\cdot\!\vX=-K$ vanishes identically, the defining
property of the class.

On flat slices the Hamiltonian constraint gives the Eulerian energy
density with no approximation in $v$,
\begin{equation}
16\pi\rho_{\rm E} \;=\; {}^{(3)}R + K^{2} - K_{ij}K^{ij}
          \;=\; -\,\bigl|\sigma(\vX)\bigr|^{2}\;\le\;0 .
\label{eq:rho}
\end{equation}
The density is nonpositive everywhere, so every point with nonzero shear
violates the weak energy condition for the Eulerian observer. Following
Lobo and Visser~\cite[Eq.~(23)]{LoboVisser:2004}, define the nonnegative
volume integral quantifier
\begin{equation}
\ELV[\vX] \;\equiv\; -\!\int_{\Sigma_t} \rho_{\rm E}\,\dd^{3}x
        \;=\; \frac{1}{16\pi}\!\int \bigl|\sigma(\vX)\bigr|^{2}\dd^{3}x .
\label{eq:ELV}
\end{equation}
All statements below concern $\ELV$ in this specified foliation. They do
not identify it with an invariant total mass or with the energy required
by a material source.

To see where this energy lives, take Nat\'ario's canonical axisymmetric
field, generated by the Stokes stream function
$\psi = v\,r^{2}f(r)\sin^{2}\theta$,
\begin{equation}
X^{r} = 2v f(r)\cos\theta,\qquad
X^{\hat\theta} = -\,v\,\bigl[\,2f(r) + r f'(r)\,\bigr]\sin\theta ,
\label{eq:canonical}
\end{equation}
where $f$ interpolates from $f\equiv 0$ inside the bubble
($r\le R$) to $f\equiv\half$ outside the wall ($r\ge R+\Delta$).
We require $f'=0$ at both faces so that the constant extension is
$H^2$ across the wall; monotonicity is not required. The interior is then
flat and the exterior is the uniform stream $\vX=v\vz$, as assumed here.

In an orthonormal spherical basis the nonzero strain components are
\begin{align}
\sigma_{rr}&=2vf'\cos\theta,&
\sigma_{\theta\theta}&=\sigma_{\phi\phi}=-vf'\cos\theta,\nonumber\\
\sigma_{r\theta}&=-v(f'+\half rf'')\sin\theta.
\end{align}
Their contraction gives the known local density~\cite{Natario:2001,Rodal:2024},
\begin{equation}
\rho_{\rm E}(r,\theta) = -\,\frac{v^{2}}{8\pi}
 \Bigl[\,3f'^{2}\cos^{2}\theta
 + \bigl(f' + \half\, r f''\bigr)^{2}\sin^{2}\theta\,\Bigr].
\label{eq:rhoclosed}
\end{equation}
We have checked this expression symbolically against an independent
evaluation of the constraint~\eqref{eq:rho}. Equation~\eqref{eq:rhoclosed} is worth
reading slowly. The two terms are the two ways the shear can point. The
first, weighted by $\cos^{2}\theta$, is the radial compression along the
axis of motion; it involves only $f'$, of order $1/\Delta$ across the
wall. The second, weighted by $\sin^{2}\theta$, is the tangential shear
in the equatorial belt; it carries the extra piece $\half rf''$, and
since $f''\sim1/\Delta^{2}$ and $r\sim R$ in the wall, this piece is of
order $R/\Delta^{2}$---larger than the first by the factor $R/\Delta$.
That single term $\half rf''$ is the return current of the
Introduction, now written down explicitly. It drives every result below.
The rest of the paper integrates it and shows that no admissible shift
can avoid it.

\section{Where the energy is}
\label{sec:where}

\begin{figure}[t]
\centering
\includegraphics[width=\linewidth]{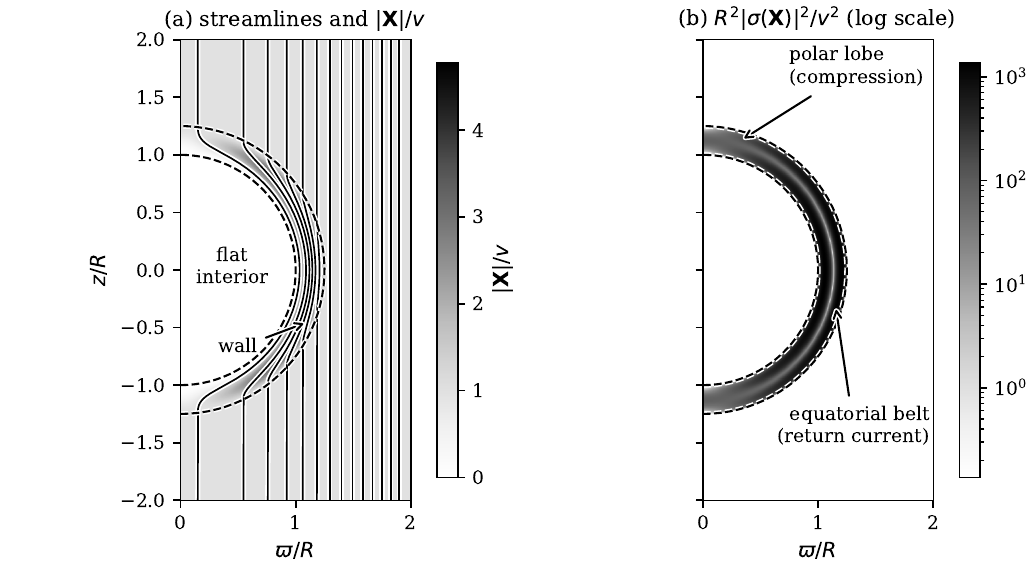}
\caption{\label{fig:mechanism}The return current and where it puts the
energy, for the canonical field~\eqref{eq:canonical} with a quintic wall
of aspect ratio $\Delta/R=0.25$ (meridional half-plane; dashed arcs mark
the wall faces $r=R$ and $r=R+\Delta$). Darker shades indicate larger
values in both panels. \emph{(a)} Streamlines, drawn at
equal flux intervals, over the shift magnitude $|\vX|/v$. The interior is
flat, the exterior is the uniform stream $v\vz$, and every streamline that
would cross the core is diverted through the wall: the displaced flux
$\pi vR^{2}$ is squeezed into a shell
of thickness $\Delta$, and the shift reaches $|\vX|\simeq4.8\,v$ inside
the wall---the return current of Eq.~\eqref{eq:return}. \emph{(b)} The
dimensionless energy density $R^{2}|\sigma|^{2}/v^{2}=-16\pi R^{2}\rho_{\rm E}/v^{2}$
on a logarithmic scale. The two
terms of Eq.~\eqref{eq:rhoclosed} are visible as distinct structures: the
polar lobes ($3f'^{2}\cos^{2}\theta$, the radial compression, peaking at
mid-wall where $f'$ is largest) and the equatorial belt
($(f'+\half rf'')^{2}\sin^{2}\theta$, the return current, peaking near
$x\simeq0.232$ and $0.805$, where $x=(r-R)/\Delta$). At this aspect ratio
the belt already carries $41$ times the integrated energy of the lobes,
and its peak density is $16$ times higher. Both ratios grow as
$(R/\Delta)^{2}$; for this quintic wall the integrated ratio approaches
$2(R/\Delta)^{2}$ exactly, the coefficient being one sixth of the profile
ratio $\int_{0}^{1}\ddot h^{2}\dd t\big/\!\int_{0}^{1}\dot h^{2}\dd t=12$
for $h(t)=6t^{5}-15t^{4}+10t^{3}$.}
\end{figure}

Integrating~\eqref{eq:rhoclosed} over all space is immediate. The
angular integrals are elementary,
$\int_{0}^{\pi}\cos^{2}\theta\sin\theta\,\dd\theta=\tfrac23$ and
$\int_{0}^{\pi}\sin^{3}\theta\,\dd\theta=\tfrac43$, and since $f'$ and
$f''$ are supported in the wall the radial integral runs only over
$[R,R+\Delta]$. One finds the exact volume integral of the canonical field,
\begin{equation}
\ELV = \underbrace{\frac{v^{2}}{2}\!\int f'^{2}\,r^{2}\dd r}_{\textstyle E_{\parallel}\ \text{(compression)}}
 + \underbrace{\frac{v^{2}}{3}\!\int\!\bigl(f'+\half rf''\bigr)^{2} r^{2}\dd r}_{\textstyle E_{\perp}\ \text{(return current)}} .
\label{eq:Esplit}
\end{equation}
The first term has the scaling used in the Lobo--Visser estimate: with
$f'\sim1/2\Delta$, $r\sim R$, and $\int_{\rm wall}\dd r\sim\Delta$,
\begin{equation}
E_{\parallel}\ \asymp\ \frac{v^{2}}{2}\,\frac{1}{4\Delta^{2}}\,R^{2}\,\Delta
 \ \asymp\ \frac{v^{2}R^{2}}{\Delta}\,,
\label{eq:Epar}
\end{equation}
recovering Eq.~\eqref{eq:LV}. The second term is governed by its
$\half rf''$ piece, with $f''\sim1/\Delta^{2}$ and $r\sim R$,
\begin{equation}
E_{\perp}\ \asymp\ \frac{v^{2}}{3}\,\frac{R^{2}}{4\Delta^{4}}\,R^{2}\,\Delta
 \ \asymp\ \frac{v^{2}R^{4}}{\Delta^{3}}\,,
\label{eq:Eperp}
\end{equation}
so that
\begin{equation}
\frac{E_{\perp}}{E_{\parallel}}\ \asymp\ \Bigl(\frac{R}{\Delta}\Bigr)^{2} .
\label{eq:ratio}
\end{equation}
Here $\asymp$ denotes the same asymptotic order, with constants depending
on the fixed rescaled profile. The tangential shear, invisible in the
diagonal of $K$, is the dominant
contribution for this canonical family with a fixed rescaled profile.
Figure~\ref{fig:mechanism} shows the mechanism directly. Panel (a) is the
flux being squeezed through the wall, and panel (b) is where that leaves
the energy, split into the polar lobes of the compression term and the
equatorial belt of the return current.

A single number makes this concrete. Take the compactly matched quintic wall
$f(r)=\half\,[\,6t^{5}-15t^{4}+10t^{3}\,]$ with $t=(r-R)/\Delta$
(so that $f$ is $C^{2}$ with $f=0$ at $r=R$ and $f=\half$ at
$r=R+\Delta$), and the parameters $v=0.1$, $R=100$, $\Delta=10$.
Evaluating~\eqref{eq:Esplit} term by term gives
\begin{equation}
E_{\parallel}=1.97,\qquad
E_{\perp}=435.4,\qquad
\ELV\simeq437.4 ,
\label{eq:numbers}
\end{equation}
in geometric units. The return-current contribution is about 221 times the
compression contribution, consistent with the asymptotic ratio
$2(R/\Delta)^{2}$ for this quintic profile. Direct three-dimensional
quadrature of the independently
computed constraint gives relative error $6.97\times10^{-8}$ on the
grid stated in Sec.~\ref{sec:numerics}.

The remaining question is whether this is special to the canonical field
\eqref{eq:canonical}. It is not: no choice of stream function can hide
the return current, because the flux that produces it is fixed by
geometry alone.

\section{The flux identity}
\label{sec:flux}

We now make the return current exact. Let $R,\Delta>0$ and $v\ge0$.
The boundary conditions are stated in the comoving frame and apply at
every aspect ratio, with the thin-wall limit taken subsequently:
\begin{quote}
\textbf{(H1)} $\vX-v\vz\in H^{1}(\mathbb{R}^{3};\mathbb{R}^{3})$ and
$\nabla\!\cdot\!\vX=0$ weakly;\\
\textbf{(H2)} $\vX\equiv 0$ on the closed ball $\bar B_{R}$;\\
\textbf{(H3)} $\vX\equiv v\vz$ on
$\mathbb{R}^{3}\setminus B_{R+\Delta}$.
\end{quote}
These say that the shift has finite Dirichlet energy relative to the
uniform exterior stream, that the interior is flat, and that the
transition happens in a shell of thickness $\Delta$. Equalities in
(H2)--(H3) are understood in the Sobolev trace sense. The canonical field
with the quintic wall is one member of this class. The arguments below
may first be read for piecewise smooth fields; the stated $H^{1}$ results
follow by density, the weak divergence theorem, and lower
semicontinuity.
For rough $H^1$ fields, $\ELV$ denotes the continuous quadratic
extension of the smooth Eulerian-density integral. No assertion about
the full stress tensor of an arbitrary rough metric is needed.

Under \emph{(H1)--(H3)}, for almost every height $|z_{0}|<R$ let
\begin{equation}
A(z_{0})=\bigl\{(x,y,z_{0}) :\,
R^{2}-z_{0}^{2}\le x^{2}+y^{2}\le (R+\Delta)^{2}-z_{0}^{2}\bigr\}
\end{equation}
be the slice of the wall at that height. Then
\begin{equation}
\int_{A(z_{0})} X_{z}\,\dd A \;=\; v\,\pi\bigl[(R+\Delta)^{2}-z_{0}^{2}\bigr],
\label{eq:fluxid}
\end{equation}
while the area $|A(z_{0})|=\pi\Delta(2R+\Delta)$ is the same at every
height.

To obtain this identity, apply the divergence theorem to $\vX$ on the piece of the wall above the
plane, $W^{+}(z_{0})=\{R\le|\mathbf{x}|\le R+\Delta,\ z\ge z_{0}\}$. Its
boundary has three parts: an outer cap on $|\mathbf{x}|=R+\Delta$ (normal
$+\rhat$), an inner cap on $|\mathbf{x}|=R$ (normal $-\rhat$), and the
annulus $A(z_{0})$ (normal $-\vz$). The inner cap contributes nothing by
(H2). On the outer cap $\vX=v\vz$ by (H3), and the flux of a uniform
field through the outer cap is, by direct integration of its normal
component, $v\pi[(R+\Delta)^{2}-z_{0}^{2}]$. This signed flux formula
also holds when $z_0<0$. Since $\nabla\!\cdot\!\vX=0$ the total
flux through $\partial W^{+}$ vanishes, and the three pieces sum to
\eqref{eq:fluxid}. The area is
$\pi[(R+\Delta)^{2}-z_{0}^{2}]-\pi[R^{2}-z_{0}^{2}]
=\pi\Delta(2R+\Delta)$, independent of $z_{0}$.

For a continuous representative the identity holds at every height; the
almost-everywhere statement is the natural one in the $H^{1}$ class and
is sufficient for the integration leading to Eq.~\eqref{eq:shiftmass}.

The average of $X_{z}$ over the annulus is therefore
\begin{equation}
\langle X_{z}\rangle_{A(z_{0})}
 = \frac{v\,[(R+\Delta)^{2}-z_{0}^{2}]}{\Delta(2R+\Delta)}
 \ \xrightarrow[\ z_{0}=0,\ \Delta\ll R\ ]{}\ \frac{vR}{2\Delta}\,,
\label{eq:return}
\end{equation}
the return current in exact form. Note what this identity does and does not
assume: it uses only that the field is divergence-free and matches the
prescribed interior and exterior. It says nothing about how the flux is
distributed within the annulus---that freedom is what the bound in the
next section removes.

\section{The lower bound}
\label{sec:bound}

For every shift satisfying \emph{(H1)--(H3)},
\begin{equation}
\ELV\;\ge\;
\frac{v^{2}R^{4}}{60\,\Delta^{3}}\,
\frac{1}{\bigl(1+\tfrac{\Delta}{2R}\bigr)\bigl(1+\tfrac{\Delta}{R}\bigr)^{2}} .
\label{eq:bound}
\end{equation}

The calculation uses three short estimates chained together. Write
$W=\{R\le|\mathbf{x}|\le R+\Delta\}$ for the wall; by (H2)--(H3) the
gradient $\nabla\vX$ is supported in $\bar W$. The first estimate turns
the flux identity into a statement about how much $\vX$ there must be;
the second, into how large its gradient must be; the third relates that
gradient to the shear that carries the energy.

First, the flux forces a minimum amount of shift:
\begin{equation}
\int_{W}|\vX|^{2}\,\dd^{3}x \;\ge\;
\frac{16\pi}{15}\,\frac{v^{2}R^{5}}{\Delta(2R+\Delta)}.
\label{eq:shiftmass}
\end{equation}
By Cauchy--Schwarz on the annulus and Eq.~\eqref{eq:fluxid},
\begin{align}
\int_{A(z_{0})}\!X_{z}^{2}\,\dd A
&\ge\frac{\bigl(\int_{A(z_{0})}X_{z}\,\dd A\bigr)^{2}}
{|A(z_{0})|}\nonumber\\
&=\frac{v^{2}\pi\,[(R+\Delta)^{2}-z_{0}^{2}]^{2}}
{\Delta(2R+\Delta)} .
\end{align}
Drop $\Delta$ against $R$ in the numerator (weakening the inequality),
integrate over $z_{0}\in(-R,R)$ with
$\int_{-R}^{R}(R^{2}-z_{0}^{2})^{2}\dd z_{0}=\tfrac{16}{15}R^{5}$, and
use $X_{z}^{2}\le|\vX|^{2}$ and $A(z_{0})\subset W$.

Second, a Poincar\'e estimate across the wall gives
\begin{equation}
\int_{W}|\vX|^{2}\,\dd^{3}x \;\le\;
\Delta^{2}\Bigl(1+\tfrac{\Delta}{R}\Bigr)^{2}\!
\int_{W}|\nabla\vX|^{2}\,\dd^{3}x.
\label{eq:poincare}
\end{equation}
Along a ray, $\vX(R\omega)=0$, so for $r\in[R,R+\Delta]$,
$|\vX(r\omega)|^{2}
 =|\!\int_{R}^{r}\partial_{\rho}\vX\,\dd\rho|^{2}
 \le\Delta\!\int_{R}^{R+\Delta}|\partial_{\rho}\vX|^{2}\dd\rho$ by
Cauchy--Schwarz. Multiply by $r^{2}\le(R+\Delta)^{2}$, integrate over $r$
and the solid angle, use $\rho^{2}\ge R^{2}$ to restore the measure on
the right, and $|\partial_{\rho}\vX|\le|\nabla\vX|$.

Third, for a solenoidal field the integrated shear and gradient satisfy
the exact identity
\begin{equation}
\int|\sigma(\vX)|^{2}\dd^{3}x
 =\half\!\int|\nabla\vX|^{2}\dd^{3}x.
\label{eq:shearidentity}
\end{equation}
Expand
$|\sigma|^{2}=\half\partial_{i}X_{j}\partial_{i}X_{j}
 +\half\partial_{i}X_{j}\partial_{j}X_{i}$ and integrate the cross term
by parts over a large ball $B_{L}\supset W$:
\begin{equation}
\int\partial_{i}X_{j}\,\partial_{j}X_{i}
=\oint_{\partial B_{L}}\! n_{i}X_{j}\partial_{j}X_{i}\,\dd A
 -\int X_{j}\,\partial_{j}(\nabla\!\cdot\!\vX)=0,
\end{equation}
since $\nabla\vX\equiv0$ on $\partial B_{L}$ and
$\nabla\!\cdot\!\vX=0$.
For an $H^1$ field apply the weak identity
$\int\partial_i u_j\partial_j u_i=\int(\nabla\cdot u)^2$
to $u=\vX-v\vz\in H^1(\mathbb R^3)$; it follows by smooth
approximation, or directly by Plancherel's theorem.

This last identity is worth pausing over: the divergence-free condition, which
created the difficulty by making $\rho_{\rm E}=-|\sigma|^{2}/16\pi$, also removes
the usual Korn constant that would otherwise enter the estimate relating shear to
gradient. No constant is lost.

Chaining Eq.~\eqref{eq:ELV} with
Eqs.~\eqref{eq:shearidentity}, \eqref{eq:poincare}, and \eqref{eq:shiftmass} in turn,
\begin{align}
16\pi\ELV
 &= \int|\sigma|^{2}
  = \half\!\int|\nabla\vX|^{2}
  \ \ge\ \frac{\int_{W}|\vX|^{2}}{2\Delta^{2}(1+\Delta/R)^{2}} \nonumber\\
 &\ge\ \frac{16\pi}{15}\,
   \frac{v^{2}R^{5}}{2\Delta^{3}(2R+\Delta)(1+\Delta/R)^{2}}\,,
\end{align}
which is Eq.~\eqref{eq:bound} on writing $2R+\Delta=2R(1+\Delta/2R)$.

Two consequences are worth stating on their own.

For $\Delta\ll R$, $\ELV\gtrsim v^{2}R^{4}/60\Delta^{3}$, larger than
the standard estimate~\eqref{eq:LV} by $\sim(R/\Delta)^{2}/60$.
For a fixed rescaled canonical profile, the estimate $v/\Delta$
captures the radial variation but misses the tangential shear.
For general admissible fields, Eq.~\eqref{eq:fluxid} forces the return
current~\eqref{eq:return} without bounding individual gradient components
from above.

Within (H1)--(H3), every admissible shift obeys a thin-wall lower
bound of order $v^{2}R^{4}/\Delta^{3}$. A fixed rescaled quintic
profile supplies an upper bound of the same order on the minimum.
Thus $\mathcal E_{\min}=\Theta(v^{2}R^{4}/\Delta^{3})$. Removing the
expansion does not bring the thin-wall Eulerian negative energy down to
the conventional volume-integral estimate $v^{2}R^{2}/\Delta$ used for
both drives. It places the zero-expansion class two powers of
$R/\Delta$ above it. At fixed nonzero $\Delta/R$, both the minimum and
that estimate scale as $v^{2}R$. The comparison is between
volume-integral quantifiers at equal $v$, $R$ and $\Delta$, and the
statement is a floor. Individual admissible shifts can carry arbitrarily
more energy.

The compact matching in (H2)--(H3) is essential to the exact flux
identity. A profile with noncompact tails requires an approximate flux
identity with explicit trace errors; the present bound should not be
applied to such a profile merely by discarding its tails.

Quantum-inequality restrictions on negative energy
densities~\cite{FordRoman:1995,Pfenning:1997wh} may impose further
constraints only after a quantum field, state, observer trajectory, and
sampling prescription have been specified. No such quantum-inequality
analysis is attempted here; Eq.~\eqref{eq:bound} is a classical geometric
statement about the chosen Eulerian volume integral.

\section{The exact optimal field at finite aspect ratio}
\label{sec:optimal}

The constant in Eq.~\eqref{eq:bound} is not optimized, so it is natural
to ask for the true minimum
$\mathcal{E}_{\min}=\min\{\ELV:\text{(H1)--(H3)}\}$.
This problem can be solved in closed form for every $\Delta/R>0$.
For $v=0$ the minimum is zero and the unique minimizer is $\vX=0$;
in deriving the normalized profile below we take $v>0$.

By the shear identity~\eqref{eq:shearidentity}, minimizing $\ELV$ is
the same as minimizing the Dirichlet energy $\int|\nabla\vX|^{2}$ over
divergence-free fields with the prescribed boundary values. The exterior
stream $v\vz$ is a pure $\ell=1$ vector harmonic, and the Dirichlet
functional does not couple different angular sectors; hence the minimizer
stays in the $\ell=1$ sector---that is, in the stream-function family
$\psi=v\,r^{2}f(r)\sin^{2}\theta$ of Eq.~\eqref{eq:canonical}. Any higher
multipole would satisfy homogeneous boundary conditions and could only
add to the energy; Appendix~\ref{app:ell1} gives the argument in detail.
It also supplies a direct Stokes calculation establishing global minimality,
without relying on angular-sector reduction.
The problem therefore collapses to the radial
functional~\eqref{eq:Esplit}, to be minimized over $f$ subject to
\begin{equation}
f(R)=0,\quad f(R+\Delta)=\tfrac12,\quad f'(R)=f'(R+\Delta)=0,
\label{eq:bcs}
\end{equation}
which are precisely the matching conditions for the $H^{1}$ shift.
Expanding Eq.~\eqref{eq:Esplit} and integrating its cross term by parts,
using $f'=0$ at both endpoints, gives the equivalent strictly convex
functional
\begin{equation}
\frac{\ELV[f]}{v^{2}}
=\frac13\int_R^{R+\Delta}r^{2}f'^{\,2}\,\dd r
+ \frac1{12}\int_R^{R+\Delta}r^{4}f''^{\,2}\,\dd r .
\label{eq:reducedfunctional}
\end{equation}
Its Euler--Lagrange equation is
\begin{equation}
r^{4}f''''+8r^{3}f'''+8r^{2}f''-8rf'=0,
\label{eq:ELfull}
\end{equation}
so the solution space in the wall is spanned by
$1,r^{2},r^{-1},r^{-3}$.

For a compact expression set $y=r/R$, $q=1+\Delta/R$,
$P=4q^{2}+7q+4$, and $S=q^{4}+q^{3}+q^{2}+q+1$. The unique solution of
Eq.~\eqref{eq:bcs} is
\begin{equation}
f_{\min}(y)=A+By^{2}+\frac{C}{y}+\frac{D}{y^{3}},
\label{eq:fexact}
\end{equation}
where
\begin{align}
A&=\frac{q(4q^{4}+4q^{3}+4q^{2}+9q+9)}
        {2(q-1)^{3}P},&
B&=-\frac{3q(q+1)}{2(q-1)^{3}P},\nonumber\\
C&=-\frac{3qS}{(q-1)^{3}P},&
D&=\frac{q^{3}(q^{2}+q+1)}{(q-1)^{3}P}.
\label{eq:coefficients}
\end{align}
Direct substitution into Eq.~\eqref{eq:reducedfunctional} yields the
finite-aspect-ratio minimum
\begin{equation}
\mathcal{E}_{\min}
=\frac{v^{2}R^{4}}{\Delta^{3}}\,c_{\min}(q),\qquad
c_{\min}(q)=\frac{3qS}{4P} .
\label{eq:exactminimum}
\end{equation}

To compare this value with an arbitrary shift satisfying \emph{(H1)--(H3)}, set
$\vY=\vX-\vX_{\min}$, where $\vX_{\min}$ is generated by
Eq.~\eqref{eq:fexact}. Then
\begin{equation}
\ELV[\vX]=\mathcal E_{\min}
 +\frac{1}{32\pi}\int_W|\nabla\vY|^2\,\dd^3x
 \ \ge\ \mathcal E_{\min}.
\label{eq:energygap}
\end{equation}
Equality holds only for $\vX=\vX_{\min}$.
The cross term vanishes by the Stokes equation for $\vX_{\min}$
and the zero trace and zero divergence of $\vY$; the explicit
calculation is given in Appendix~\ref{app:ell1}. This also establishes
existence and uniqueness over the full three-dimensional class.

The minimum has a classical hydrodynamic meaning. Set
$\lambda=R/(R+\Delta)$. The concentric-sphere
Stokes resistance factor is~\cite[Eq.~(33)]{Nangia:2017}
\begin{equation}
\mathcal K(\lambda)=
\frac{1-\lambda^5}{1-\frac94\lambda+\frac52\lambda^3
 -\frac94\lambda^5+\lambda^6}.
\end{equation}
For viscosity $\mu$, the dissipation and energy minimum are
\begin{align}
\mathcal D&=2\mu\int|\sigma|^2
             =6\pi\mu Rv^2\mathcal K,\\
\mathcal E_{\min}&=\frac{\mathcal D}{32\pi\mu}
             =\frac{3Rv^2\mathcal K}{16}.
\end{align}
The latter reduces exactly to Eq.~\eqref{eq:exactminimum}.
The Stokes solution itself is
classical~\cite{Nangia:2017}. Its relativistic reading is the new
element. The viscous dissipation of creeping flow between concentric
spheres, divided by $32\pi\mu$, is the least Eulerian negative energy
that a zero-expansion bubble with these boundary data can carry. The
Stokes pressure enters as the Lagrange multiplier that enforces zero
divergence, not as a material pressure.

\begin{figure}[t]
\centering
\includegraphics[width=\linewidth]{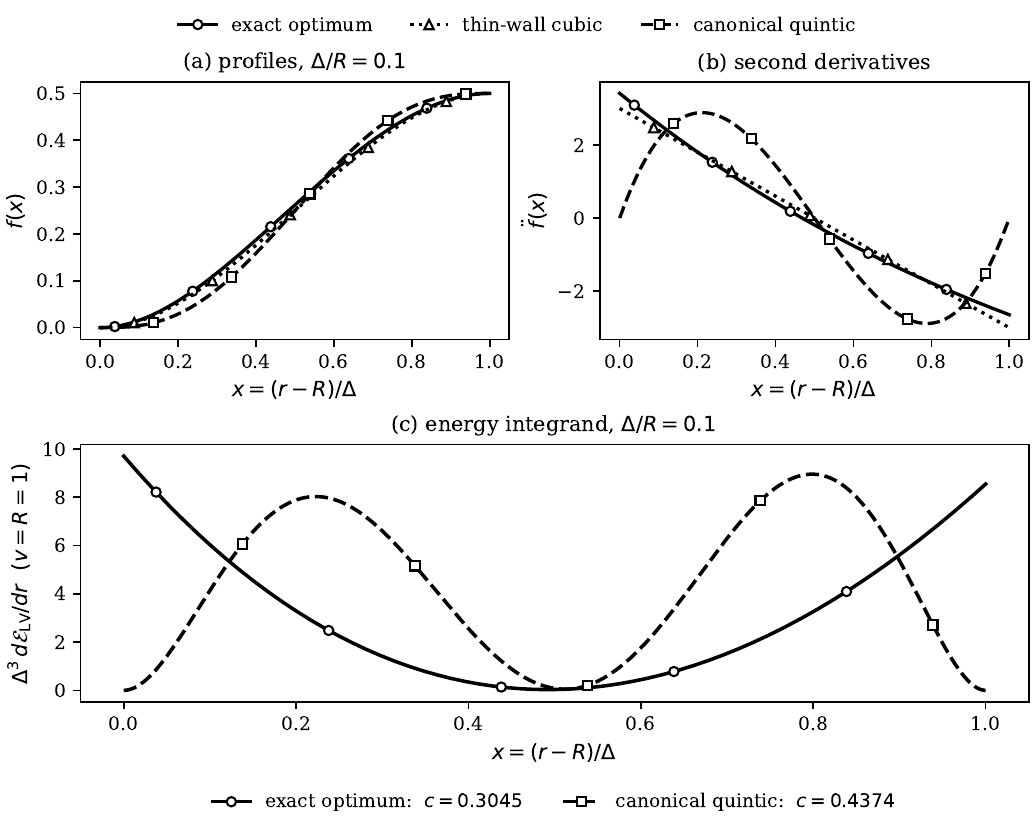}
\caption{\label{fig:optimal}The exact optimal wall profile.
\emph{(a)} The finite-aspect solution~\eqref{eq:fexact} at
$\Delta/R=0.1$ (solid, circles), its cubic thin-wall limit (dotted,
triangles), and the canonical quintic (dashed, squares).
\emph{(b)} Their second derivatives with respect to $x$, using the
same line styles and markers. \emph{(c)} The radial integrands
of the exact optimum and the quintic:
$c_{\min}=0.3045$, while the quintic gives $c=0.4374$, a factor $1.44$
larger.}
\end{figure}

The thin-wall result follows rather than being assumed. With
$a=\Delta/R$,
\begin{equation}
c_{\min}=\frac14+\frac12a+\frac{13}{30}a^{2}+O(a^{3}),
\label{eq:cseries}
\end{equation}
and, in $x=(r-R)/\Delta$,
\begin{equation}
f_{\min}(x)\longrightarrow f^{*}(x)=\tfrac32x^{2}-x^{3}.
\label{eq:cubic}
\end{equation}
Thus $\mathcal{E}_{\min}\sim v^{2}R^{4}/(4\Delta^{3})$ with a sharp
leading constant $1/4$. The canonical quintic has thin-wall coefficient
$5/14$, exactly $10/7$ times the optimum.
Figure~\ref{fig:optimal} compares the three profiles, their second
derivatives, and the radial energy integrands of the exact optimum and
the quintic. Table~\ref{tab:optimal} collects the
coefficients at representative aspect ratios. The exact minimum energy lies
below the cubic's energy at every finite ratio, the two converging as the wall
thins. At the tabulated aspect ratios the canonical quintic stays above
both. All three tabulated energies remain more
than an order of magnitude above the bound~\eqref{eq:bound}, whose
constants are not optimized.

The exact optimizer belongs to $H^{2}(R,R+\Delta)$ as a radial profile, and the
associated shift belongs to the $H^{1}$ class (H1). In general $f''$ does
not match the zero exterior value at the wall faces, so the extended
profile need not be $C^{2}$. Smooth compactly matched profiles are dense
in the same affine energy space and can round the two faces in arbitrarily
thin sublayers; their values of $\ELV$ converge to
Eq.~\eqref{eq:exactminimum}. Hence Eq.~\eqref{eq:exactminimum} is the
minimum in the natural Sobolev class and the infimum in the smooth
subclass.
For example, subtract a smooth compactly matched reference profile
from $f_{\min}$. The difference belongs to $H^2_0(R,R+\Delta)$ and
can be approximated by compactly supported smooth functions in the
$H^2$ norm. Adding back the reference preserves both boundary values
and derivatives, and continuity of Eq.~\eqref{eq:reducedfunctional}
gives the stated infimum. This energy-space result does not assert
that the limiting nonsmooth metric has a regular material source.

\section{Numerical verification}
\label{sec:numerics}

The argument is a chain of identities and inequalities, and we checked
each link against an independent evaluation of the
constraint~\eqref{eq:rho}. We used the canonical field~\eqref{eq:canonical}
with the quintic wall of Sec.~\ref{sec:where} (and, as a control, an
independent $C^{3}$ septic profile), and computed the shear and its
integrals symbolically. For $v=0.1$, $R=100$, $\Delta=10$:

\begin{itemize}
\item[(i)] \emph{Flux identity~\eqref{eq:fluxid}} holds to relative
error below $3\times10^{-11}$ (quadrature-limited) at heights
$z_{0}/R=0,\,0.4,\,0.8$ at the stated parameter point.
\item[(ii)] \emph{Return current~\eqref{eq:return}:} the mean vertical
shift over the equatorial annulus is $0.576190$, matching
$v(R+\Delta)^{2}/[\Delta(2R+\Delta)]=0.576190$ ($\simeq vR/2\Delta$).
\item[(iii)] \emph{Shear identity~\eqref{eq:shearidentity}:}
$\int|\sigma|^{2}/\int|\nabla\vX|^{2}=0.49999998$, with the cross term
divided by the gradient integral equal to $-3.49\times10^{-8}$.
\item[(iv)] \emph{Two-term split~\eqref{eq:Esplit}:}
$E_{\parallel}=1.97$, $E_{\perp}=435.4$, reproducing the closed
form~\eqref{eq:numbers}. The analytically evaluated one-dimensional value,
$\ELV\simeq437.3912338$, agrees with the independent three-dimensional
quadrature to relative error $7\times10^{-8}$ on a $3000\times400$ grid
in $(r,\theta)$.
\item[(v)] \emph{The bound~\eqref{eq:bound}} holds throughout, with
margins $27$--$51$ (Table~\ref{tab:bound}), consistent with the
unoptimized constants---the sharp one-dimensional Poincar\'e constant
alone would improve $60\to60\cdot4/\pi^{2}\approx24$.
\item[(vi)] \emph{Scaling.} Doubling $v$ multiplies $\ELV$ by
$4.000$; doubling $R$ by $14.4$ (approaching $16$ as $\Delta/R\to0$);
doubling $\Delta$ by $0.154$ (approaching $0.125$). At
$\Delta/R=10^{-2}$ the exact ratios are $15.8404$ and $0.12754$,
checking the canonical quintic's thin-wall scaling without treating finite
ratios as fitted exponents.
\item[(vii)] \emph{Exact minimum~\eqref{eq:exactminimum}.} Substitution
of Eq.~\eqref{eq:fexact} satisfies the boundary conditions and differential
equation exactly, and its functional equals Eq.~\eqref{eq:exactminimum}.
A Ritz calculation with
eight homogeneous basis functions independently reproduces
$c_{\min}$ with relative error below $10^{-8}$ for the aspect ratios in
Table~\ref{tab:optimal}. Three-dimensional quadrature for the cubic gives
$c=0.25252$ at $\Delta/R=5\times10^{-3}$, consistent with its exact
one-dimensional value $0.252518$ and the limit $1/4$.
\end{itemize}

\begin{table}[t]
\caption{\label{tab:optimal}The coefficient
$c=\ELV\Delta^{3}/v^{2}R^{4}$ for the exact minimum, the cubic
thin-wall profile, and the canonical quintic, compared with the lower
bound~\eqref{eq:bound}, at three wall aspect ratios; and the thin-wall
limit. Values are rounded to four decimal places.}
\centering
\begin{tabular}{lcccc}
\toprule
$\Delta/R$ & $c_{\min}$ & $c_{\rm cubic}$ & $c_{\rm canonical}$ & bound\\
\midrule
$0.30$ & $0.4440$ & $0.4760$ & $0.6601$ & $0.0086$\\
$0.10$ & $0.3045$ & $0.3075$ & $0.4374$ & $0.0131$\\
$0.02$ & $0.2602$ & $0.2603$ & $0.3718$ & $0.0159$\\
$\to0$ & $0.2500$ & $0.2500$ & $0.3571$ & $0.0167$\\
\bottomrule
\end{tabular}
\end{table}

\begin{table}[t]
\caption{\label{tab:bound}Exact Eulerian volume integral versus the
bound~\eqref{eq:bound} (geometric units), for the canonical field with
two wall profiles. Entries are rounded to the shown precision; the margin
is the ratio of the exact integral to the bound.}
\centering
\begin{tabular}{lcccc}
\toprule
profile & $(v,R,\Delta)$ & $\ELV$ & bound & margin\\
\midrule
quintic & $(0.1,100,10)$ & $4.37\times10^{2}$ & $1.31\times10^{1}$ & 33 \\
quintic & $(0.1,200,10)$ & $6.32\times10^{3}$ & $2.36\times10^{2}$ & 27 \\
quintic & $(0.1,100,20)$ & $6.72\times10^{1}$ & $1.32\times10^{0}$ & 51 \\
quintic & $(0.2,100,10)$ & $1.75\times10^{3}$ & $5.25\times10^{1}$ & 33 \\
septic  & $(0.1,100,10)$ & $6.48\times10^{2}$ & $1.31\times10^{1}$ & 49 \\
\bottomrule
\end{tabular}
\end{table}

\section{Discussion}
\label{sec:discussion}

The central result is a sharp statement about the whole zero-expansion
class. Every metric of the form~\eqref{eq:metric} whose shift
satisfies (H1)--(H3) carries at least the
Eulerian negative energy $\mathcal E_{\min}$ of
Eq.~\eqref{eq:exactminimum}, and
$\mathcal E_{\min}\sim v^{2}R^{4}/(4\Delta^{3})$ as the wall thins.
The minimum is attained, so the constant $1/4$ cannot be improved, and
the energy-gap identity~\eqref{eq:energygap} measures exactly how far
any other admissible shift lies above it. The result is a floor, not a
ceiling. Adding any nonzero divergence-free field $t\vY$ with zero wall
traces to the minimizer adds $t^2\int|\nabla\vY|^2/(32\pi)$ to the
energy, which can be arbitrarily large.

\emph{Physical scope.} The bound holds for the unit-lapse, flat-slice,
stationary, exactly divergence-free class with compact matching. Within
that class, local Eulerian WEC violation is unavoidable and its volume
integral has the floor established here. The result does not specify an
equation of state or a microphysical source, and $\ELV$ should not be
added to a payload mass or identified with ADM energy.

\emph{The optimal field.} Section~\ref{sec:optimal} identifies the
minimum-energy field exactly at every finite aspect ratio. The
minimal-curvature cubic is its thin-wall limit, not an ansatz. The
canonical quintic lies $43\%$ above the optimum asymptotically. Changing the
rescaled wall profile can alter the leading coefficient, but no admissible
shift can lie below the sharp minimum.

\emph{Curvature diagnostics.} For $v=c$, bubble radius $5\,\mathrm{m}$,
and inverse wall thickness $4\,\mathrm{m}^{-1}$, Rodal~\cite{Rodal:2024}
reports curvature-invariant amplitudes about 35 times larger for Nat\'ario
than for Alcubierre. The large $rf''$ term identified here may contribute
to enhanced Einstein-sector curvature in thin walls. Rodal's pointwise
invariant comparison includes Weyl-sector curvature, whereas $\rho_{\rm E}$
is determined by the Einstein-sector constraint. A universal scaling
statement for every curvature invariant would require a separate
invariant-by-invariant analysis.

\emph{Positive-energy shells.} The bound concerns the zero-expansion
warp class, where the negative Eulerian density is structural. It says
nothing about
the positive-energy shell constructions
of~\cite{BobrickMartire:2021,Fuchs:2024}. Those positive-energy examples
lie outside the combined flat-slice, unit-lapse, zero-expansion
hypotheses used to derive Eq.~\eqref{eq:rho}, so the
present bound does not apply to them.

The return current is kinematic. It follows from incompressibility and
the boundary data alone, so it sets an unavoidable scale
$v^{2}R^{4}/\Delta^{3}$ for the Eulerian negative-energy integral of the
compactly matched zero-expansion class, and the Stokes field realizes
that scale exactly. Natural extensions are matching with controlled
noncompact tails, nonunit lapse, and the corresponding floors for
individual curvature invariants.

\begin{acknowledgements}
During preparation of this manuscript, OpenAI Codex (GPT-5) and Anthropic
Claude Opus 5 assisted with technical review of derivations and claims and
with drafting and editorial revision. Nelson Bol\'ivar directed and checked
this assistance. The authors reviewed the final manuscript and are
responsible for its contents.
\end{acknowledgements}

\section*{Data availability}
The code and numerical data supporting the findings of this study are
available from the authors to editors, referees, and readers upon
reasonable request.

\section*{Conflict of interest}
The authors declare that they have no conflict of interest.

\appendix

\section{Why the minimizer is purely \texorpdfstring{$\ell=1$}{l=1}}
\label{app:ell1}

We justify the reduction used in Sec.~\ref{sec:optimal}: among
divergence-free fields obeying (H1)--(H3), the minimizer of the Dirichlet
energy lies in the stream-function family~\eqref{eq:canonical}.

Write the minimization as follows. Fix any admissible reference field
$\vX_{0}$ (for instance the canonical field with a quintic wall). Every
other admissible field is $\vX=\vX_{0}+\vY$ with
\begin{equation}
\nabla\!\cdot\!\vY=0, \qquad
\vY\big|_{|x|\le R}=0, \qquad \vY\big|_{|x|\ge R+\Delta}=0 ,
\label{eq:Ycond}
\end{equation}
that is, $\vY$ is a solenoidal $H^{1}$ field supported in the wall $W$
with zero trace on both wall faces. Finiteness of the Dirichlet energy
does not impose a zero normal derivative. Denote this space by
$\mathcal{V}$.
Because the Dirichlet functional is quadratic,
\begin{equation}
\int|\nabla(\vX_{0}+\vY)|^{2}
 = \int|\nabla\vX_{0}|^{2}
 + 2\!\int\! \partial_{i}X_{0j}\,\partial_{i}Y_{j}
 + \int|\nabla\vY|^{2}.
\label{eq:quadratic}
\end{equation}

Now expand $\vY$ in vector spherical harmonics on each sphere
$|x|=r$. The three families---the radial harmonics
$Y_{\ell m}\rhat$ and the two tangential families
$r\nabla Y_{\ell m}$ and $\mathbf{r}\times\nabla Y_{\ell m}$---are
mutually orthogonal in $L^{2}(S^{2})$, and the flat-space Laplacian acts
within each $(\ell,m)$ sector without mixing them. Consequently the
Dirichlet form is block-diagonal:
\begin{equation}
\int|\nabla\vY|^{2}\,\dd^{3}x
 \;=\; \sum_{\ell,m}\ \int|\nabla\vY_{\ell m}|^{2}\,\dd^{3}x ,
\label{eq:blockdiag}
\end{equation}
and likewise the cross term in~\eqref{eq:quadratic} pairs only components
with the same $(\ell,m)$. The reference field $\vX_{0}$ is axisymmetric
and purely $\ell=1$: its angular dependence is $\cos\theta$ in the radial
component and $\sin\theta$ in the polar one, i.e.\ $Y_{10}$ and
$\partial_{\theta}Y_{10}$ (the toroidal family is absent because the
field is poloidal, being derived from a Stokes stream function).
Therefore the cross term vanishes for every component of $\vY$ with
$(\ell,m)\neq(1,0)$, and~\eqref{eq:quadratic} becomes
\begin{align}
\int|\nabla\vX|^{2}
 &=\int|\nabla(\vX_{0}+\vY^{(10)})|^{2}\nonumber\\
 &\quad+\sum_{(\ell,m)\neq(1,0)}
       \int|\nabla\vY_{\ell m}|^{2}.
\label{eq:split}
\end{align}
The last sum is a sum of nonnegative terms, none of which is required by
the boundary data (each $\vY_{\ell m}$ obeys homogeneous conditions).
Setting them to zero can only lower the energy, so the minimizer has
$\vY_{\ell m}=0$ for $(\ell,m)\neq(1,0)$ and remains axisymmetric and
$\ell=1$.

Within $(\ell,m)=(1,0)$ the toroidal component is likewise orthogonal to
the poloidal reference field in the Dirichlet form, obeys homogeneous
boundary conditions, and contributes a nonnegative term. It therefore
vanishes at the minimum. Finally, within the remaining $\ell=1$ poloidal
sector every divergence-free field
admits a Stokes stream function, $\psi = v\,g(r)\sin^{2}\theta$ with
$g=r^{2}f$; substituting this into the strain gives
Eq.~\eqref{eq:rhoclosed} and hence the radial
functional~\eqref{eq:Esplit}. This completes the reduction. We note that
the argument uses only the flat background geometry and the fact that the
boundary data $v\vz$ is a single vector harmonic. General exterior
data can require additional sectors.

A direct Stokes calculation gives the same result. Write
$F(r)=f_{\min}(r/R)$ and define on $W$
\begin{align}
p(r,\theta)&=vH(r)\cos\theta,\nonumber\\
H(r)&=r^2F'''+6rF''+4F'.
\end{align}
The vector Laplacian of Eq.~\eqref{eq:canonical} has components
\begin{align}
(\nabla^2\vX_{\min})_r
 &=2v(F''+4F'/r)\cos\theta,\nonumber\\
(\nabla^2\vX_{\min})_{\hat\theta}
 &=-v(rF'''+6F''+4F'/r)\sin\theta.
\end{align}
Equation~\eqref{eq:ELfull} gives
$H'=2(F''+4F'/r)$, hence
$\nabla^2\vX_{\min}=\nabla p$ on $W$.
For every $\vY\in\mathcal V$, integration by parts yields
\begin{align}
\int_W\nabla\vX_{\min}:\nabla\vY
 &=-\int_W\nabla p\cdot\vY\nonumber\\
 &=\int_W p\,\nabla\cdot\vY=0.
\end{align}
The zero trace removes both boundary terms. The identity holds for
$H^1_0$ variations by weak integration by parts. Expanding the
Dirichlet energy and using Eq.~\eqref{eq:shearidentity} gives
Eq.~\eqref{eq:energygap}. Its remainder vanishes only if $\vY=0$,
by the zero trace. This calculation includes nonaxisymmetric and
toroidal variations.

\end{document}